\documentclass{article}
\usepackage{amsmath,amsthm,amsfonts,amssymb}
\usepackage{microtype}
\usepackage{graphicx}
\usepackage{url}
\usepackage{dsfont}

\DeclareMathOperator*{\Exp}{\mathbb{E}}
\DeclareMathOperator*{\Var}{\mathrm{Var}}

\DeclareMathAlphabet{\varmathbb}{U}{bbold}{m}{n}
\newcommand{\one}{\mathds{1}}

\begin{document}

\theoremstyle{definition}
\newtheorem{definition}{Definition}
\newtheorem{theorem}{Theorem}
\newtheorem{lemma}{Lemma}
\newtheorem{corollary}{Corollary}
\newtheorem{claim}{Claim}
\newtheorem{proposition}{Proposition}
\newtheorem{exercise}{Exercise}

\newcommand{\Z}{{\mathbb{Z}}}
\newcommand{\R}{{\mathbb{R}}}
\newcommand{\F}{{\mathbb{F}}}
\newcommand{\eps}{\varepsilon}
\renewcommand{\emptyset}{\varnothing}
\newcommand{\e}{\mathrm{e}}
\newcommand{\poly}{\mathrm{poly}}
\newcommand{\inner}[2]{\langle #1, #2 \rangle}
\newcommand{\norm}[1]{\| #1 \|}
\newcommand{\bnorm}[1]{\big\| #1 \big\|}
\newcommand{\Bnorm}[1]{\Big\| #1 \Big\|}

\newcommand{\abs}[1]{\left| #1 \right|}
\renewcommand{\P}{\textsf{P}}
\newcommand{\BPP}{\textsf{BPP}}
\newcommand{\TIME}{\textsf{TIME}}
\newcommand{\SPACE}{\textsf{SPACE}}

\renewcommand{\v}{\mathbf{v}}
\renewcommand{\u}{\mathbf{u}}
\newcommand{\z}{\mathbf{z}}
\newcommand{\vpar}{\v^{\parallel}}
\newcommand{\vperp}{\v^{\perp}}
\newcommand{\zpar}{\z^{\parallel}}
\newcommand{\zperp}{\z^{\perp}}

\title{Optimal $\eps$-biased sets with\\ just a little randomness}
\author{Cristopher Moore\thanks{\texttt{moore@santafe.edu}, Santa Fe Institute and Department of Computer Science, University of New Mexico} 
\and Alexander Russell\thanks{\texttt{acr@cse.uconn.edu}, Department of Computer Science and Engineering, University of Connecticut}} 
\maketitle

\begin{abstract}
Subsets of $\F_2^n$ that are \emph{$\eps$-biased}, meaning that the parity of any set of bits is even or odd with probability $\eps$ close to $1/2$, 
are powerful tools for derandomization.
 A simple randomized construction shows that such sets exist of size $O(n/\eps^2)$, and known deterministic constructions achieve sets of size $O(n/\eps^3)$, $O(n^2/\eps^2)$, and $O((n/\eps^2)^{5/4})$.  Rather than derandomizing these sets completely in exchange for making them larger, we attempt a partial derandomization while keeping them small, constructing sets of size $O(n/\eps^2)$ with as few random bits as possible.  The naive randomized construction requires $O(n^2/\eps^2)$ random bits.  We give two constructions.  The first uses Nisan's space-bounded pseudorandom generator to partly derandomize a folklore probabilistic construction of an error-correcting code, and requires $O(n \log (1/\eps))$ bits.  
Our second construction requires $O(n \log (n/\eps))$ bits, but is more elementary; it adds randomness to a Legendre symbol construction on Alon, Goldreich, H{\aa}stad, and Peralta, and uses Weil sums to bound high moments of the bias.
\end{abstract}

\section{Introduction}

Derandomization is the art of replacing random choices with deterministic ones.  In many cases, we can accomplish this by finding explicit constructions of combinatorial objects that ``look random'' in some sense.  In particular, say a set $S \subseteq \F_2^n$ fools a function $f$ if
\[
\abs{ \Exp_{x \in S} f(x) - \Exp_{x \in \F_2^n} f(x) } \le \eps \, ,
\]
for some small $\eps$.  If there are families of sets of polynomial size that we can construct in polynomial time, such that for each constant $c$ we can fool every $f \in \TIME(n^c)$ with $\eps$ sufficiently small, then every polynomial-time randomized algorithm can be derandomized and $\P=\BPP$.

In a number of applications, even sets that fool linear functions are useful~\cite{naor-naor}.  In $\F_2^n$, any such function is the parity of some subset of $x$'s coordinates.  Let $x \in \F_2^n$ and let $T \subseteq [n]$.  The parity of the bits of $x$ indexed by $T$ is
\[
f_T(x) = \sum_{i \in T} x_i \, . 
\]
We say that $S$ is \emph{$\eps$-biased} if, for all $T \ne \emptyset$, 
\[
\abs{ \Pr_{x \in S} \left[ f_T(x) = 0 \right] - \Pr_{x \in S} \left[ f_T(x) = 1 \right] } \le \eps \, .
\]
Equivalently, if we identify $\F_2^n$ with $\{\pm 1\}^n$ in the natural way, then 
\[
\abs{ \Exp_{x \in S} \phi_T(x) } \le \eps 
\quad \text{where} \quad
\phi_T(x) = \prod_{i \in T} x_i \, . 
\]
This is the same as saying that $\chi_S$, the characteristic function of $S$, has a nearly flat Fourier spectrum: if we normalize it to $(1/|S|) \chi_S$, it has  no coefficients greater than $\eps$ in absolute value.  As a consequence, sampling a function on $S$ gives a good approximation of its expectation if its Fourier spectrum has bounded $\ell_1$ norm.  In addition, $\eps$-biased sets are important building blocks in other pseudorandom constructions; for instance, if $\eps = n^{-c}$ for $c > 0$, then an $\eps$-biased set is also approximately $O(\log n)$-wise independent.

There is a nice duality between $\eps$-based sets and linear error-correcting codes~\cite{azar}.  Given an $\eps$-biased set $S$, the truth table of each parity function $f_T(x)$ is a string in $\F_2^{|S|}$.  Each such string is nearly balanced, with Hamming weight between $(1-\eps) |S|/2$ and $(1+\eps) |S|/2$.  The set of parity functions has rank $n$, so an $\eps$-biased set $S \in \F_2^n$ yields a $(|S|, n, d)$ code, i.e., a code of length $|S|$, rank $n$, and distance $d=(1-\eps)|S|/2$.  As a consequence, we can lower bound the size of an $\eps$-biased set using sphere-packing arguments.  As long as $\eps$ is not too small, and in particular if $\eps = 1/\poly(n)$, this gives $|S| = \Omega(n/(\eps^2 \log \eps^{-1}))$~\cite{aghp}.

This lower bound is essentially tight, since we can construct an $\eps$-biased set by choosing $O(n / \eps^2)$ elements of $\F_2^n$ uniformly and independently.  Equivalently, a random error-correcting code meets the Gilbert-Varshamov bound with high probability.  Of course, this requires $n|S| = O(n^2/\eps^2)$ random bits.  Under the reasonable assumption that $\TIME(2^{O(n)}) \not\subseteq \SPACE(2^{O(n)})$, this construction can be generically derandomized~\cite{cheraghchi}.  But, as always, we are interested in derandomized constructions that work even in the absence of complexity assumptions.

Starting with~\cite{naor-naor}, several deterministic constructions have been discovered, yielding $\eps$-biased sets of size polynomial in $n$ and $1/\eps$.  Depending on how $\eps$ scales with $n$, the best known constructions~\cite{aghp,ben-aroya-ta-shma} yield sets of size $O(n/\eps^3)$, $O(n^2/\eps^2)$, and $O((n/\eps^2)^{5/4})$.  The construction of~\cite{ben-aroya-ta-shma} is especially notable; it applies Bezout's theorem from algebraic geometry, and achieves a set whose size is the $5/4$ power of the optimum.

Tradeoffs between randomness and the quality of a combinatorial object is a classic topic in theoretical computer science.  Here, we explore a different part of the randomness-size plane.  Rather than reducing the amount of randomness to zero at the cost of making the set larger, we ask how much randomness we need to construct a set of optimal size, or equivalently what spaces we can succinctly describe that are guaranteed to contain at least one $\eps$-biased set of optimal size.  

Specifically, we give two randomness-efficient constructions of $\eps$-biased sets of size $O(n/\eps^2)$.  While neither construction offers a witness that the resulting set is indeed $\epsilon$-biased, both succeed with probability arbitrarily close to $1$.  The first uses $O(n \log (1/\eps))$ random bits, using Nisan's space-bounded generator to partly derandomize a construction of random error-correcting codes.  The second uses $O(n \log (n/\eps))$ bits but is more elementary, and has a pleasant algebraic flavor: it works by ``re-randomizing'' a construction in~\cite{aghp} involving the Legendre symbol, and we use Weil sums to bound high moments of the bias.  Note that if $\eps = n^{-c}$ for $c > 0$, then in both constructions the number of random bits we need is much smaller than the set itself.  

\section{A random error-correcting code and Nisan's generator}

First we review a simple folklore construction of a random linear error-correcting code whose distance is very close to half its length; this corresponds to an $\eps$-biased set using the duality mentioned above.  This construction already does noticeably better than the naive one, using $O(|S|)=O(n/\eps^2)$ random bits.  We then derandomize it further using a standard space-bounded pseudorandom generator, reducing the number of random bits to $O(n \log (1/\eps))$.

Let $m \ge n$ and consider the finite field $\F_{2^m}$.  We can identify each $x \in \F_2^n$ with an element of $\F_{2^m}$ in a way that preserves the additive structure of $\F_2^n$ by setting all but the last $n$ bits to zero.  For any fixed $\alpha \in \F_m$, we can then define a set of codewords in $\F_2^n \times \F_{2^m}$, 
\[
C_\alpha = \left\{ w_x = \big( x, \alpha x \big) \mid x \in \F_2^n \right\} \, .
\]
Since multiplication by $\alpha$ is a linear function, $C_\alpha$ is closed under addition, making it a linear code.  It has rank $n$ and length $n+m$.  We will show that, if $\alpha \in \F_{2^m}$ is uniformly random and $m/n$ is sufficiently large, then $C_\alpha$ has distance $(1-\eps)(n+m)/2$ with high probability, in which case it corresponds to an $\eps$-biased set of size $n+m$.  Equivalently, the Hamming weight $|w_x| = |x| + |\alpha x|$ of every nonzero codeword is at least $(1-\eps)(n+m)/2$.

For each nonzero $x \in \F_2^n$, $\alpha x$ is uniformly random in $\F_{2^m}$ since $\alpha$ is.  Let $\delta \le 1/2$.  By the union bound, the probability that there is an $x \ne 0$ such that $|w_x| \le \delta (n+m)$ is at most
\[
P
= 2^{-m} \sum_{k,j: k+j \le \delta (n+m)} {n \choose k} {m \choose j} \, ,
\]
where we sum over $k=|x|$ and $j=|\alpha x|$.  This sum has at most $n^2$ terms, and the summand is maximized when $k=\delta n$ and $j = \delta m$, so
\begin{equation}
\label{eq:sump}
P 
\le n^2 \,2^{-m} {n \choose \delta n} {m \choose \delta m} 
\le n^2 \,2^{-m} \,\e^{h(\delta) (n+m)} \, , 
\end{equation}
where $h(\delta) = -\delta \ln \delta - (1-\delta) \ln (1-\delta)$ denotes the entropy function.  

Now if $\delta = (1-\eps)/2$, the Taylor series gives
\[
h(\delta) \le \ln 2 - \frac{\eps^2}{2} \, , 
\]
and~\eqref{eq:sump} becomes
\[
P \le n^2 \,\e^{n \ln 2 - (\eps^2/2) (n+m)} \, . 
\]
If we set 
\[
m = A \frac{n}{\eps^2} 
\] 
for some constant $A > 2 \ln 2$, then $P=2^{-\Omega(n)}$.  Thus $C_\alpha$ has distance $(1-\eps)(n+m)/2$ with high probability, giving an $\eps$-biased set of size $n+m = O(n/\eps^2)$.  

To choose $\alpha$ uniformly would take $m = O(n/\eps^2)$ random bits.  However, we can do better by applying a pseudorandom generator for space-bounded computation.  First, let us modify the construction somewhat, using $t = m/n = O(1/\eps^2)$ blocks of $n$ bits each.  Rather than choosing $\alpha$ from $\F_{2^m}$, we write $\alpha=(\alpha_1, \ldots, \alpha_t)$ where $\alpha_i \in \F_{2^n}$ for each $i$.  We then define each codeword as a concatenation of $t+1$ blocks, 
\[
C_\alpha = \left\{ w_x = \big(x, \alpha_1 x, \alpha_2 x, \ldots, \alpha_t x \big) \mid x \in \F_{2^n} \right\} \, . 
\]
If the $\alpha_i$ are uniformly random in $\F_{2^n}$ then so is $\alpha_i x$, and the probability that any $w_x$ has Hamming weight less than $(1-\eps)(n+m)/2$ is $2^{-\Omega(n)}$ just as before.

Now note that, for each $x \in \F_{2^n}$, there is a branching program $B_x$ with states $\{0, \ldots, n+m\}$ that takes $\alpha_1, \ldots, \alpha_t$ as input and computes the total Hamming weight of $w_x$.  Its initial state is $|x|$, on the $i$th step it reads $\alpha_i$ and increments its state by the weight of $\alpha_i x$, and it accepts if $|w_x| \ge (1-\eps)(n+m)/2$.  Our goal is to fool $B_x$ with a pseudorandom sequence of $tn$ bits, in such a way that the probability distribution of its final state has a total variation distance $o(2^{-n})$ from the distribution induced by uniformly random $\alpha_i$.  Taking a union bound over all $x$, the probability that any $B_x$ rejects, i.e., that any $w_x$ has Hamming weight less than $(1-\eps)(n+m)/2$, is then $o(1)$ just as if the $\alpha_i$ were uniform.  In that case, $C_\alpha$ is again an error-correcting code of distance $(1-\eps)(n+m)/2$ with high probability.

We do this with Nisan's pseudorandom generator for space-bounded computation.  Say that $f:\{0,1\}^\ell \to \{0,1\}^{bt}$ is a \emph{pseudorandom generator for block size $b$ and space $s$ with parameter $\delta$ and seed length $\ell$} if, for all branching programs $B$ that read $b$ bits at each step, take $t$ steps, and have width at most $2^s$, 
\[
\abs{ 
\Pr_{\gamma \in \{0,1\}^\ell} \big[ \mbox{$B(f(\gamma))$ accepts}\, ] 
- \Pr_{\alpha \in \{0,1\}^{bt}} \big[ \mbox{$B(\alpha)$ accepts}\, ] 
} \le \delta \, . 
\]
Then Lemma 3 of~\cite{nisan} states the following, in slightly different notation:
\begin{lemma}
Let $t \le 2^{n/20}$.  Then there is an explicit pseudorandom generator $f$ for block size $b$ and space $b/20$ with parameter $2^{-b/20}$ and seed length $O(b \log t)$.
\end{lemma}

In our case, to match the union bound over all $2^n$ possible $x$ we want the parameter $\delta$ to be, say, $2^{-2n}$.  To this end, we modify $B_x$ so that it reads $b=40 n$ bits at each step, ignoring all but $n$ of them.  Then~\cite{nisan} gives a pseudorandom generator with seed length $O(n \log t) = O(n \log (1/\eps))$.  

Indeed, the space our branching program needs is just $\log (n+m+1) = O(\log (n/\eps))$, far smaller than the $b/20=\Theta(n)$ allowed by the lemma.  Moreover, if we think of $B_x$ as computing $|w_x| \bmod (n+m+1)$ (note that $|w_x|$ will never actually wrap around) it becomes a permutation branching program.  Furthermore, for uniform inputs we know the probability distribution on the program's states exactly, namely the binomial distribution.  

It is tempting to think that these facts allow us to reduce the randomness still further, say to $O(n + \log (1/\eps))$.  However, to our knowledge even the best known derandomization results on branching programs under various assumptions~\cite{raz-reingold,braverman,de,steinke} require $\Omega((\log 1/\delta)(\log t))$ random bits, even for constant width.  Since $\delta = 2^{-\Omega(n)}$ and $t = \Omega(1/\eps^2)$, this again gives $\Omega(n \log (1/\eps))$ random bits.

\section{A construction using the Legendre symbol and Weil sums}

Here we present another construction, which uses $O(n \log (n/\eps))$ random bits.  If $\eps$ is fairly large, say $\eps = 1/n^{o(1)}$, then this uses $O(\log n)$ more randomness than the previous construction.  However, it is elementary and extremely explicit, and lets us invoke some pretty algebra.

First we recall the definition of the Legendre symbol.  Given a prime $q$, let $g$ be a primitive root, i.e., a multiplicative generator of $\F_q^\times$.  Then let $\chi: \F_q \rightarrow \R$ be defined as follows:
\[
\chi(x) = \begin{cases} 
+1 & \mbox{if $x=z^2$ for some $z \ne 0$} \\
-1 & \mbox{if $x=gz^2$ for some $z \ne 0$} \\
0 & \mbox{if $x=0$} \, ,
\end{cases} 
\]
This is the quadratic multiplicative character of $\F_q^\times$, extended to $\F_q$ by setting $\chi(0)=0$.  Thus $\chi(xy)=\chi(x) \chi(y)$ for all $x, y \in \F_q$.  

Alon, Goldreich, H{\aa}stad, and Peralta~\cite{aghp} used the Legendre symbol to construct an $\eps$-biased set as follows.  For each $x \in \F_q$, consider the sequence 
\[
w(x) = \left( \chi(x+1), \chi(x+2), \ldots, \chi(x+n) \right) \, . 
\]
Mapping $\{\pm 1\}$ to $\{0,1\}$ gives an element $\F_2^n$; if $x+i=0$, we define $w(x)_i=1$.  Their set is then 
\[
S = \{ w(x) \mid x \in \F_q \} \, . 
\]
Except for a small error due to the rare case where $x+i=0$, the bias of $S$ with respect to $T \subseteq [n]$ is then
\[
b_T 
= \Exp_{x \in \F_q} \phi_T(w(x))
= \Exp_{x \in \F_q} \prod_{i \in T} w(x)_i
= \Exp_{x \in \F_q} \prod_{i \in T} \chi(x+i) 
= \Exp_{x \in \F_q} \chi\!\left( \prod_{i \in T} (x+i) \right) \, .
\]
If we write
\[
p(x) = \prod_{i \in T} (x+i) \, , 
\]
then $p(x)$ is a polynomial of degree $|T| \le n$.  In that case, the bias is a Weil sum, which we can bound using the following classic theorem:
\begin{theorem}[Weil]
\label{thm:weil}
Let $p(x) \in \F_q[x]$ be a non-square polynomial of degree $d$. Then
$$
\abs{ \Exp_{x \in \F_q} \chi(p(x))} \leq \frac{d-1}{\sqrt{q}} \, .
$$
\end{theorem}
\noindent
Since $d \le n$, the bias is bounded by $|b_T| \le n/\sqrt{q}$.  This gives an $\eps$-biased set $S$ of size $q = n^2 / \eps^2$.

Our approach is to ``re-randomize'' this construction.  Rather than taking $n$ consecutive Legendre symbols, we let $\Sigma = (s_1, \ldots, s_n) \in F_q^n$ be a collection of $n$ ``shifts.''  For each $x \in \F_q$, these shifts let us extract $n$ bits from the Legendre symbol sequence, giving a string 
$$
w(x) = \left( \chi(x + s_1), \chi(x + s_2), \ldots, \chi(x + s_n) \right) \, .
$$
In return for choosing these shifts randomly, we get to use a field $\F_q$ considerably larger than the set itself.  We then show that $S$ is $\eps$-biased with high probability in $\Sigma$ by using Theorem~\ref{thm:weil} to control high moments of the bias.

Let $X \subseteq \F_q$ be an arbitrary set of size $\ell$, such as $\{1,\ldots,\ell\}$.  Letting $x$ range over $X$ yields a set
\[
S = \{ w(x) \mid x \in X \} \subseteq \F_2^n \, , 
\]
with $|S|=\ell$.  Assume for now that $x+s_j \ne 0$ for all $x \in X$ and all $j \in [n]$.  
Then the bias $S$ with respect to $T \subseteq [n]$ is
$$
b_T 
= \abs{ \Exp_{x \in X} \phi_T(w(x)) }
= \abs{ \Exp_{x \in X} \prod_{j \in T} \chi(x - s_j) } \, .
$$
We will show that, with high probability in $\Sigma$, this bias is small for all $T \ne \emptyset$.  To this end, we bound its $2k$th moment for some $k$ to be determined below.  Expanding its $2k$th power gives products of the form 
\begin{equation}
\label{eq:prod}
\prod_{t=1}^{2k} \prod_{j \in T} \chi(x_t - s_j) \, , 
\end{equation}
averaged over all tuples $\{x_1,\ldots,x_{2k}\} \in X^{2k}$.  For each $x \in X$, let $N(x)$ be the number of times that $x$ appears in this tuple.  If $N(x)$ is even for all $x$, then this product is a square, and is $1$ regardless of the $s_j$.  Taking the union bound over all $(2k-1)!! = (2k-1)(2k-3) \cdots 3 \cdot 1$ perfect matchings of $2k$ objects, the probability that this occurs---given that all $\ell^{2k}$ tuples $\{x_1,\ldots,x_{2k}\}$ are equally likely---is at most 
\[
\frac{(2k-1)!!}{\ell^k}
= \frac{(2k)!}{2^k k! \,\ell^k} 
\le \sqrt{2} \left( \frac{2k}{\e \ell} \right)^{\!k} \, ,
\]
where we used a form of Stirling's inequality.

On the other hand, if $N(x)$ is odd for some $x \in X$, the product~\eqref{eq:prod} can be written 
\[
\prod_{j \in T} \chi\!\left( p_{x_1,\ldots,x_{2k}}(s_j) \right) \, , 
\]
where 
\[
p_{x_1,\ldots,x_{2k}}(s) = \prod_{x: \text{$N(x)$ odd} } (x-s)
\]
is a polynomial of degree at most $2k$.  In that case, since the $s_j$ are independent and uniform in $\F_q$, Theorem~\ref{thm:weil} gives
\[
\abs{ \Exp_\Sigma \prod_{j \in T} \chi\!\left( p_{x_1,\ldots,x_{2k}}(s_j) \right) }
= \abs{ \Exp_{s \in \F_q} \chi\!\left( p_{x_1,\ldots,x_{2k}}(s) \right) }^{|T|}
\le \left( \frac{2k-1}{\sqrt{q}} \right)^{\!|T|} \, . 
\]

Putting this all together, we have
\begin{align}
\Exp_\Sigma b_T^{2k}
&= \Exp_\Sigma \left( \Exp_{x \in X} \prod_{j \in T} \chi(x_i - s_j) \right)^{\!2k\,} \nonumber \\
&= \Exp_{x_1, \ldots, x_{2k}} \Exp_\Sigma \prod_{t=1}^{2k} \prod_{j \in T} \chi(x_t - s_j) \nonumber \\
&\le \Pr_{\{x_1,\ldots,x_{2k}\}}\!\left[ \text{$N(x)$ even for all $x$}\right] \nonumber \\
&\quad + \Exp_{x_1, \ldots, x_{2k}} \left[ \Exp_\Sigma \prod_{t=1}^{2k} \prod_{j \in T} \chi(x_{t}-s_j) \;\Bigg\vert\; \text{$N(x)$ odd for some $x$} \right] \nonumber \\
&\le \sqrt{2} \left( \frac{2k}{\e \ell} \right)^{\!k} + \left(\frac{2k}{\sqrt{q}}\right)^{\!|T|} \, . 
\label{eq:moment}
\end{align}
Markov's inequality gives
\begin{equation}
\label{eq:markov}
\Pr\!\left[ |b_T| > \eps \right] = \Pr[b_T^{2k} > \eps^{2k}] \le \frac{\Exp_\Sigma b_T^{2k}}{\eps^{2k}}
\end{equation}
We now set $q = 4 (\e \ell)^2$, making the field quadratically larger than $|S| = \ell$.  We also set $k=|T|$, using the $2k$th moment to control parities of weight $k$.  Then combining~\eqref{eq:moment} and~\eqref{eq:markov} gives, for any $|T| \ge 1$,
\[
\Pr\!\left[ |b_T| > \eps \right] \le 2 \left( \frac{2|T|}{\e \ell \eps^2} \right)^{\!|T|} \, . 
\]
Taking a union bound over all $T \ne \emptyset$ and using ${n \choose |T|} \le (\e n/|T|)^{|T|}$, the probability that any nontrivial parity has bias greater than $\eps$ is at most
\begin{equation}
\label{eq:geom}
\sum_{T \ne \emptyset} \Pr\!\left[ |b_T| > \eps \right] 
\le 2 \sum_{|T|=1}^n {n \choose |T|} \left( \frac{2|T|}{\e \ell \eps^2} \right)^{\!|T|} 
\le 2 \sum_{|T|=1}^n  \left( \frac{2n}{\ell \eps^2} \right)^{\!|T|} \, . 
\end{equation}
If we set 
\[
\ell = \frac{6 n}{\delta \eps^2} \, ,
\]
where $\delta \le 1$, then bounding~\eqref{eq:geom} with a geometric series gives
\[
\sum_{T \ne \emptyset} \Pr\!\left[ |b_T| > \eps \right] 
\le \frac{2\delta/3}{1-\delta/3} \le \delta \, ,
\]
so the set $S$ is $\eps$-biased with probability $1-\delta$.  Finally, our assumption that $x+s_j \ne 0$ for all $x \in X$ and all $j \in [n]$ holds with probability $1-n \ell / q = 1-O(\delta \eps^2)$.

How much randomness do we need for this construction?  We have to select the shifts $s_1,\ldots,s_n$ independently and uniformly from $\F_q$, and 
\[
q = 4 (\e \ell)^2 = O\!\left( \frac{n^2}{\eps^4} \right) \, . 
\]
Thus the number of random bits we need is
\[
n \log q =  O\big( n \log (n/\eps) \big) \, .
\]

\section{Further derandomization?}

Can we do better?  Our approach has a natural barrier at $n$ random bits; since we take a union bound over all $2^n$ index sets $T$, we need a probability space of size at least $2^n$.  Thus any further derandomization, say to $o(n)$ random bits, would have to bound the bias for many parities simultaneously.

The situation is similar for constructing optimal Ramsey graphs, i.e., edge-colored complete graphs on $n$ vertices such that the largest monochromatic clique has size less than $k=2 \log n$.  As pointed out in~\cite{naor-ramsey}, we can do this by choosing the coloring from a ${k \choose 2}$-wise $\eps$-biased distribution, i.e., a family of functions from the set of edges to $\{0,1\}$ such that the parity of any set of ${k \choose 2}$ or fewer edges is odd with probability $\eps$-close to $1/2$.  If $\eps = 2^{-k^2}$, the probability that a given clique of size $k$ is monochromatic is $o\big(1/{n \choose k}\big)$.  So, by the union bound, with high probability there are no monochromatic cliques of size $k$.  

We can generate such families~\cite{aghp} with $O\big( \log \log n + {k \choose 2} +\log \eps^{-1} \big) = O(\log^2 n)$ random bits.  Since we need a probability space of size ${n \choose k} = 2^{\Omega(\log^2 n)}$ for the union bound over all ${n \choose k}$ cliques to work, this is tight---unless we can do better than the union bound, ensuring simultaneously that many cliques are bichromatic.  It is an interesting open question whether this can be reduced to, say, $O(\log n)$ random bits, in which case there are explicit graph families of polynomial size guaranteed to consist largely of optimal Ramsey graphs.

\paragraph{Acknowledgments.}   We thank Avi Wigderson and Mark Braverman for helpful conversations, and Micha{\l} Kotowski for catching several typos.  This work was supported by NSF grant CCF-1117426 and ARO contract W911NF-04-R-0009.  We are also grateful to the Scholium Project, and particularly the Androkteinos, for inspiration.

\end{document}